\documentclass{ws-ijmpa}

\begin{document}

\markboth{Gen Nakamura and Kazuhiro Yamamoto}
{Quantum Larmor radiation from a moving charge
in an electromagnetic plane wave background}

%
\catchline{}{}{}{}{}
%

\title{Quantum Larmor radiation from a moving charge
in an electromagnetic plane wave background}

\author{GEN NAKAMURA}

\address{
Department of Physical Science, Hiroshima University,\\
Higashi-Hiroshima 739-8526,~Japan}

\author{KAZUHIRO YAMAMOTO}
\address{
Department of Physical Science, Hiroshima University,\\
Higashi-Hiroshima 739-8526,~Japan\\
kazuhiro@hiroshima-u.ac.jp
}

\maketitle

\begin{history}
\end{history}

\begin{abstract}
We extend our previous work [Phys.~Rev.~D{\bf 83}~045030~(2011)], which investigated the
first-order quantum effect in the Larmor radiation from a moving charge in a spatially 
homogeneous time-dependent electric field. 
Specifically, we investigate the quantum Larmor radiation from a moving charge 
in a monochromatic electromagnetic plane wave background based on the scalar quantum 
electrodynamics at the lowest order of the perturbation theory. Using the in-in formalism,
we derive the theoretical formula of the total radiation energy from a charged particle 
in the initial states being at rest and being in a relativistic motion.
Expanding the theoretical formula in terms of the Planck constant $\hbar$, we obtain 
the first-order quantum effect on the Larmor radiation.
The quantum effect generally suppresses the total radiation energy compared with the 
prediction of the classical Larmor formula, which is a contrast to the previous work.
The reason is explained by the fact that the radiation from a moving charge 
in a monochromatic electromagnetic plane wave is expressed in terms of the inelastic
collisions between an electron and photons of the background electromagnetic waves. 
\end{abstract}

\ccode{PACS numbers: 41.60.-m, 12.20.-m, 04.62.+v}

\def\bfp{{\bf p}}
\def\bfk{{\bf k}}
\def\bfx{{\bf x}}
\def\bfy{{\bf y}}
\def\bfV{{\bf V}}
\def\bfz{{z}}
\def\rmi{{\rm i}}
\section{Introduction}
A series of the previous works\cite{Higuchi,YN,KNY,NSY} focused the investigation 
on the quantum effect on the Larmor radiation from a charge in an accelerated 
motion on a non-trivial background field. 
These investigations were partially motivated by a series of Higuchi and Martin's 
work\cite{HM1,HM2,HM3}, which derived the radiation reaction force on a charged particle 
in an accelerated motion on the basis of the quantum electrodynamics (QED). 
The Larmor radiation from a moving charge in an accelerated motion is well-known 
in the classical electrodynamics\cite{Jackson}, and Higuchi and Martin showed how it
can be reproduced in the limit of $\hbar\rightarrow0$, where $\hbar$ is the Planck constant,  
in the framework of the scalar quantum electrodynamics (SQED).

Higuchi and Martin's works led the study on the quantum correction to the classical Larmor 
radiation using their theoretical framework. 
In Ref.\refcite{Higuchi},  Higuchi and Walker investigated the quantum correction to 
the classical Larmor radiation from a moving charge whose motion is non-relativistic.
They found that the quantum correction is non-local and that the quantum effect 
suppresses the total radiation when the particle is in a non-relativistic motion.
In our previous work\cite{YN}, we extended the work by Higuchi and Walker to the 
case of a moving charge in a relativistic motion, assuming a spatially homogeneous 
time-dependent electric field background. We found that the quantum effect 
may increase the total radiation energy compared with the classical evaluation 
when the particle is in a relativistic motion\cite{YN}. 

In the previous works, however, it has not been clearly understood what determines the 
decrease or increase of the total radiation energy of the quantum Larmor radiation.
The results in the previous works might rely on the assumption of the spatially 
homogeneous electric field background.
Further investigation in more general background might give us a hint to understand the 
origin of the quantum correction to the Larmor radiation. 
Then, as an extension of the previous works, we consider the case 
of a monochromatic electromagnetic plane wave background in the present paper.  
We evaluate the radiation energy in the framework of the SQED using the solution for
the complex scalar field equation on a monochromatic electromagnetic plane wave, 
which is a counterpart of the Volkov solution for the Dirac equation\cite{Volkov}.

There have been many works on the radiation process from a moving charge in an 
electromagnetic plane wave background so far (e.g., Ref.\refcite{NikishovRitus,BK,Brown}).
However, our work in the present paper is different from the previous works 
in the following points. First, we derive the theoretical formula for the radiation 
energy starting with the theoretical framework of the in-in formalism\cite{Wein}. 
Second, we demonstrate that the quantum effect generally suppresses the total radiation 
energy by explicitly evaluating the contribution at the order of $\hbar$. 
We also demonstrate that the quantum effect of the order of $\hbar$ is
explained {\it only} by that the radiation process is expressed in terms of the inelastic 
collisions between an electron and photons of the monochromatic electromagnetic wave 
background. 

The reason why we should use the in-in formalism is explained as follows. 
The in-in formalism is useful to avoid the subtle problem of the definition 
of the vacuum state for charged particles.
For example, when we consider a charged particle in a constant electric field, 
we should choose the different vacuum state for the initial state at the infinitely
past time and the final state at infinitely future time\cite{Schwinger}.
Then, the radiation process due to interaction from the charged particle must 
be carefully treated\cite{Nikishov}. 
The in-in formalism formulated by Weinberg\cite{Wein}, which was developed to evaluate higher 
order correlation functions in an inflationary universe, is useful to treat this 
problem\cite{KNY,YN}.

It is known that a charged particle moving in a curved space-time generally gives rise 
to radiation, and the classical radiation reaction force was first derived by 
De~Witt and Brehme\cite{DWB}, and Hobbs\cite{Hobbsa}. 
The radiation reaction force in the conformally flat universe is also investigated in Ref.\refcite{Hobbsb}. 
Then, several authors investigated the quantum effect on the radiation 
from a moving charged particle in an expanding universe\cite{Futamase,NSY,HWU}. 
It is demonstrated that the in-in formalism is useful to investigate the quantum effect on 
the radiation from a moving charge in an expanding universe\cite{KNY}, because the 
vacuum state for charged particles is changed there.
The authors in Ref.\refcite{KNY} demonstrated that the quantum effect suppresses the total radiation 
energy in comparison with the classical counterpart (see also~Ref.\refcite{NSY}).

This paper is organized as follows. In section 2, we derive a general formula 
for the quantum radiation from a charged particle 
in a monochromatic electromagnetic plane wave background based on 
the SQED. 
We consider the initial states that the incident charged particle is at 
rest and in a relativistic motion along with the direction of the 
polarization vector.
In section 3, we present the explicit formulas for the radiation 
with these initial states.
Then, we discuss about the origin of the quantum effect by extracting 
the terms of the order of $\hbar$.
Section 4 is devoted to summary and conclusions. 
Throughout this paper, we adopt the unit in which the speed of light equals unity $c=1$, 
and the metric convention $(-,+,+,+)$.

\section{formulation}
In this section, we derive the formula for the radiation energy 
from a moving charge in an electromagnetic plane wave 
background.
We consider the scalar QED with the action,
\begin{eqnarray}
 S=\int d^4x\left[-\eta^{\mu\nu}
\left(\partial_\mu - \frac{ie}{\hbar}A_\mu\right)\phi^\dagger
\left(\partial_\nu + \frac{ie}{\hbar}A_\nu\right)\phi
-\frac{m^2}{\hbar^2}\phi^\dagger\phi
-\frac{1}{4\mu_0}F^{\mu\nu}F_{\mu\nu}\right],
\end{eqnarray}
where $e$ and $m$ are the charge and mass of the complex scalar field $\phi$,
respectively, $\eta^{\mu\nu}$ is the metric of the Minkowski space-time, 
$A^\mu$ is the 
electromagnetic field, the field strength is defined by 
$F^{\mu\nu}=\partial^\mu A^\nu-\partial^\nu A^\mu$,
and $\mu_0$ is the magnetic permeability of vacuum.
We work in the Minkowski space-time, but consider the 
electromagnetic plane wave background.

The equation of motion of the complex scalar field is written
\begin{eqnarray}
 \left[\left(\partial_\mu+\frac{ie}{\hbar}\bar{A}_\mu\right)
\left(\partial^\mu+\frac{ie}{\hbar}\bar{A}^\mu\right)
-\frac{m^2}{\hbar^2}\right]\phi{=0},
\end{eqnarray}
where $\bar{A}^\mu$ is the background electromagnetic field.
The quantized field is written as
\begin{eqnarray}
 \phi=\sum_{\bf p}\sqrt{\frac{\hbar}{V}}(\phi_{\bf p}b_{\bf p}+
\phi^\ast_{\bf p}c^\dagger_{\bf p}),
\end{eqnarray}
where $V$ is the volume of a box, $b_{\bf p}$ and  $c^\dagger_{\bf p}$ 
are the annihilation operator of the particle and the creation
operator of the anti-particle, respectively, and $\phi_{\bf p}$ is 
the mode function decomposed in the finite box.
These operators satisfy the commutation relation,
$[b_{\bf p},b^\dagger_{{\bf p}^\prime}]=\delta_{{\bf p},{\bf p}^\prime}$,
$[c_{\bf p},c^\dagger_{{\bf p}^\prime}]=\delta_{{\bf p},{\bf p}^\prime}$,
and the other combinations are zero, 
and the vacuum state is defined by
$ b_{\bf p}\left|0\right\rangle_{b_{\bf p}}=0$, and 
$c_{\bf p}\left|0\right\rangle_{c_{\bf p}}=0$.
We assume that the background electromagnetic field depends 
only on $k\cdot x$, where the dot denotes the four dimensional 
contraction, $k\cdot x\equiv k^\mu x_\mu$. We adopt the Lorentz 
gauge condition, i.e., $\partial^\mu \bar{A}_\mu=0$, where 
$\bar{A}_\mu=\bar{A}_\mu(k\cdot x)$.
In this situation, we can solve the equation of motion as 
follows (cf. Ref.\refcite{Volkov}). 
First, we rewrite the equation of motion as
\begin{eqnarray}
 \left(\partial_\mu\partial^\mu+
2\frac{ie}{\hbar}\bar{A}_\mu\partial^\mu
-\frac{e^2}{\hbar^2}\bar{A}^2-\frac{m^2}{\hbar^2}\right)\phi_\bfp=0,
\end{eqnarray}
where $\bar{A}^2 \equiv \bar{A}\cdot\bar{A}$.
We seek the solution of the form,
\begin{eqnarray}
 \phi_\bfp=\sqrt{\frac{\hbar}{2p^0}}e^{ip\cdot x/\hbar}g_p(k\cdot x),
\end{eqnarray}
with the initial condition,
 $\phi_\bfp\sim\sqrt{\frac{\hbar}{2p^0}}e^{ip\cdot x/\hbar}$, 
at $t\rightarrow -\infty$.
Using the relations $k^2=0$, $\partial^\mu \bar{A}_\mu=0$
and $p^2=-m^2$, equation for $g_p(k\cdot x)$ becomes
\begin{eqnarray}
 {dg_p(k\cdot x)\over d(k\cdot x)}+\frac{i}{2\hbar p\cdot k}
\left(2ep\cdot \bar{A}+e^2\bar{A}^2\right)g_p(k\cdot x)=0.
\end{eqnarray}
Then, we find the solution
\begin{eqnarray}\label{phip}
  \phi_\bfp=\sqrt{\frac{\hbar}{2p^0}}e^{ip\cdot x/\hbar}\exp\left[-i\int^{k\cdot x}_{-\infty}
\frac{1}{2\hbar p\cdot k}[2ep\cdot \bar{A}((k\cdot x)^\prime)
+e^2\bar{A}^2((k\cdot x)^\prime)]d(k\cdot x)^\prime\right].
\nonumber\\
\end{eqnarray}

On the other hand, the quantized fluctuations of the electromagnetic field 
can be described
\begin{eqnarray}
 A_\mu=\sqrt{\frac{\mu_0\hbar}{V}}\sum_{(\lambda),{\bf k}^\prime}
\epsilon^{(\lambda)}_\mu \sqrt{\frac{1}{2k^\prime}}
e^{i{\bf k}^\prime\cdot{\bf x}}
(a^{(\lambda)}_{{\bf k}^\prime}e^{-ikt}
+a^{\dagger (\lambda)}_{-{\bf k}^\prime}e^{ikt}),
\end{eqnarray}
where $\epsilon^{(\lambda)}_\mu$ is the polarization vector,
$a^{\dagger(\lambda)}_{{\bf k}^\prime}$ and $a^{(\lambda)}_{{\bf k}^\prime}$
are the creation and annihilation operator of the electromagnetic 
field, respectively, which satisfy the commutation relation
$ [a^{(\lambda)}_{{\bf k}},a^{\dagger(\lambda^\prime)}_{{\bf k}^\prime}]
=\delta_{(\lambda),(\lambda^\prime)}\delta_{{\bf k},{\bf k}^\prime}$,~ 
$ [a^{(\lambda)}_{{\bf k}},a^{(\lambda^\prime)}_{{\bf k}^\prime}]=
 [a^{\dagger(\lambda)}_{{\bf k}},a^{\dagger(\lambda^\prime)}_{{\bf k}^\prime}]=0$
and the vacuum state is defined by
$ a^{(\lambda)}_{\bf k}\left|0\right\rangle_{{a}_{\bf k}}=0.
$

\begin{figure}[pb]
\centerline{\psfig{file=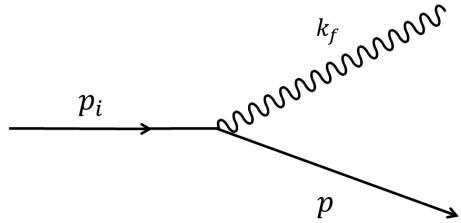,width=6.7cm}}
\vspace*{8pt}
\caption{Feynman diagram for the process. \label{fig:one} }
\end{figure}

We consider the lowest-order contribution of the process
so that one photon is emitted from a charged particle with
the initial momentum $\bfp_i$, 
as shown in Fig. \ref{fig:one}, whose interaction Hamiltonian 
is 
\begin{eqnarray}\label{interactionH}
H_I=-\frac{ie}{\hbar}\int d^3{\bf x}A^\mu
\left\{(\partial_\mu\phi^\dagger)\phi-\phi^\dagger(\partial_\mu\phi)
-\frac{2ie}{\hbar}\bar{A}_\mu\phi^\dagger\phi\right\}.
\end{eqnarray}
Substituting Eq.~(\ref{phip}) into Eq.~(\ref{interactionH}), we obtain
\begin{eqnarray}
:H_I:&=&-e\int d^3{\bf x}
\sqrt{\frac{\mu_0}{4\hbar V^3}}\sum_{(\lambda),{\bf k}^\prime,{\bf p}_1,{\bf p}_2}
\epsilon^{(\lambda)\mu}\frac{1}{\sqrt{2k^\prime p_1^0 p_2^0}}
e^{i{\bf k}^\prime\cdot{\bf x}}
(a^{(\lambda)}_{{\bf k}^\prime}e^{-ikt}
+a^{\dagger (\lambda)}_{-{\bf k}^\prime}e^{ikt})
\nonumber \\
&\times&\left(I_{p_1\mu}+I_{p_2\mu
	 }+2e\bar{A}_\mu\right)
g^\ast_{p_1}g_{p_2}e^{-i(p_1-p_2)\cdot x/\hbar}b^\dagger_{\bfp_1}b_{\bfp_2},
\end{eqnarray}
where we defined 
\begin{eqnarray}
I_{p\mu}\equiv p_\mu-\frac{k_\mu}{2p\cdot k}
(2ep\cdot\bar{A}+e^2\bar{A}^2).
\end{eqnarray}
In order to derive the total radiation energy,
we first evaluate the expectation value of the number 
operator\cite{KNY,YN}.
Using the in-in formalism\cite{Wein,AEL}, we may 
write the expectation value of the number operator
of the emitted photon with the wavenumber $\bfk_f$,
up to the second-order of the perturbative expansion,
as follows,
\begin{eqnarray}
\sum_{(\lambda)} \left\langle a^{\dagger(\lambda)}_{\bfk_f}
a^{(\lambda)}_{\bfk_f}\right\rangle
&=&\sum_{(\lambda)}\frac{1}{\hbar^2}{\rm Re}
\left[\int^\infty_{-\infty}dt_2\int^{\infty}_{-\infty}dt_1
 \left\langle
  H_I(t_1)a^{\dagger(\lambda)}_{\bfk_f}a^{(\lambda)}_{\bfk_f}H_I(t_2)
 \right\rangle\right], \nonumber\\
&=&{\rm Re}\Bigl[\frac{\mu_0e^2}{8\hbar V^3}
\int^\infty_{-\infty}dt_2\int^{\infty}_{-\infty}dt_1 
\int d^3{\bf x}_1\int d^3{\bf x}_2\nonumber\\
&&~~
\sum_{(\lambda),(\lambda_1),\bfk_1,(\lambda_2),\bfk_2}
\sum_{\bfp_1,\bfp_2,\bfp_3,\bfp_4}
\epsilon^{(\lambda_1)}_\mu\epsilon^{(\lambda_2)}_{\nu}
\frac{1}{\sqrt{k_1 k_2 p^0_1 p^0_2 p^0_3 p^0_4}}\nonumber \\
&\times&\left\langle{\rm in}\right|
(a^{(\lambda_1)}_{\bfk_1}e^{-ikt_1}+a^{\dagger(\lambda_1)}_{-\bfk_1}e^{ikt_1})
a^{\dagger(\lambda)}_{\bfk_f}a^{(\lambda)}_{\bfk_f}
\nonumber\\
&&~~~~~~
(a^{(\lambda_2)}_{\bfk_2}e^{-ikt_2}+a^{\dagger(\lambda_2)}_{-\bfk_2}e^{ikt_2})
b^{\dagger}_{\bfp_1}b_{\bfp_2}b^{\dagger}_{\bfp_3}b_{\bfp_4}
\left|{\rm in}\right\rangle
\nonumber \\
&\times&
e^{-i(\bfp_1-\bfp_2-\hbar \bfk_1)\cdot \bfx_1/\hbar}
e^{-i(\bfp_3-\bfp_4-\hbar \bfk_2)\cdot \bfx_2/\hbar}
e^{i(p_{10}-p_{20})t_1/\hbar}
e^{i(p_{30}-p_{40})t_2/\hbar}
\nonumber \\
&\times& 
(I^\mu_{p_1}+I^\mu_{p_2}+2e\bar{A}^\mu)_{x_1}
(I^\nu_{p_3}+I^\nu_{p_4}+2e\bar{A}^\nu)_{x_2}
(g^\ast_{p_1}g_{p_2})_{x_1}
(g^\ast_{p_3}g_{p_4})_{x_2}\Bigr].
\nonumber\\
\end{eqnarray}
We adopt the initial state as the one-particle state with the momentum $\bfp_i$, 
which is described by
$ \left|{\rm in}\right\rangle=
\left|0\right\rangle_{a^{(\lambda)}_\bfk}\otimes b^\dagger_{\bfp_{\rm
i}}\left|0\right\rangle_{b_\bfp}\otimes\left| 0\right\rangle_{c_\bfp}$,
then we have
\begin{eqnarray}
 \sum_{(\lambda)} \left\langle a^{\dagger(\lambda)}_{\bfk_f}
a^{(\lambda)}_{\bfk_f} \right\rangle
&=&{\rm Re}\Bigl[\frac{\mu_0e^2}{8\hbar V^3}
\int^\infty_{-\infty}dt_2\int^{\infty}_{-\infty}dt_1 
\int d^3{\bf x}_1\int d^3{\bf x}_2\
\nonumber\\
&&\sum_{\bfp}\frac{1}{k_fp^0_{\rmi}p^0}
e^{-i(p_{\rmi}-p-\hbar k_f)\cdot x_1/\hbar}
e^{-i(p-p_{\rmi}+\hbar k_f)\cdot x_2/\hbar}\nonumber \\
&&\times
(I^\mu_{p_{\rmi}}+I^\mu_{p}+2e\bar{A}^\mu)_{x_1}
(I_{p_{\rmi}\mu}+I_{p\mu}+2e\bar{A}_\mu)_{x_2}
(g^\ast_{p_{\rmi}}g_{p})_{x_1}
(g^\ast_{p}g_{p_{\rmi}})_{x_2}\Bigr].\nonumber\\
\label{expect}
\end{eqnarray}
Note that $g^\ast_{p_{\rmi}}g_p$ gives the phase factor, 
\begin{eqnarray}\label{gtimesg}
 g^\ast_{p_{\rmi}}g_p=
\exp\left[\frac{i}{\hbar}\int d(k\cdot x)
\left\{e\left(\frac{p_\rmi\cdot\bar{A}}{p_\rmi\cdot k}-
\frac{p\cdot\bar{A}}{p\cdot k}\right)+\frac{e^2\bar{A}^2}{2}
\left(\frac{1}{p_\rmi\cdot k}-\frac{1}{p\cdot k}\right)\right\}\right].
\end{eqnarray}
The vector potential of the monochromatic electromagnetic plane 
wave is written 
\begin{eqnarray}
 \bar{A}_\mu=\frac{(a_\mu e^{ik\cdot x}+
a^\ast_\mu e^{-ik\cdot x})}{2},
\end{eqnarray}
for general arbitrary polarization, where $a_\mu$ is a complex 
constant vector to specify the amplitude and the polarization.
We introduce dimensionless parameters as
\begin{eqnarray}
&&\sigma_{\rmi p}= \frac{e^2}{4}a\cdot a^\ast
\left(\frac{1}{p_\rmi\cdot k}-\frac{1}{p\cdot k}\right),\\
&&\xi_{\rmi p}e^{i\alpha_{\rmi p}}= e
\left(\frac{p_\rmi\cdot a}{p_\rmi\cdot k}-
\frac{p\cdot a}{p\cdot k}\right),\\
&&\eta_{\rmi p}e^{i\beta_{\rmi p}}
=\frac{e^2}{4}a\cdot a
\left(\frac{1}{p_\rmi\cdot k}-\frac{1}{p\cdot k}\right).
\end{eqnarray}
Note that $\eta_{\rmi p}=0$ for the circular polarization
while $\eta_{\rmi p}=\sigma_{\rmi p}, \alpha_{\rmi p}=\beta_{\rmi p}=0$
for the linear polarization. We also introduce 
$\sigma_{p \rmi}(=-\sigma_{\rmi p})$, 
$\xi_{p \rmi}(=-\xi_{\rmi p})$, 
$\eta_{p \rmi}(=-\eta_{\rmi p})$,
$\alpha_{p\rmi}(=\alpha_{\rmi p})$, and  
$\beta_{p\rmi}(=\beta_{\rmi p})$. Then, Eq.~(\ref{gtimesg}) reduces to
\begin{eqnarray}
  g^\ast_{p_{\rmi}}g_p&=&
\exp\left[\frac{i}{\hbar}\int d(k\cdot x)
\left(\sigma_{\rmi p}+\xi_{\rmi p}\cos(k\cdot x+\alpha_{\rmi p})
+\eta_{\rmi p}\cos(2k\cdot x +\beta_{\rmi p})\right)\right]
\nonumber
\\
&=&
\exp\left[\frac{i}{\hbar}\left(\sigma_{\rmi p}k\cdot x
+\xi_{\rmi p}\sin(k\cdot x+\alpha_{\rmi p})+\frac{\eta_{\rmi p}}{2}
\sin(2k\cdot x+\beta_{\rmi p})\right)\right],
\label{eqgg}
\end{eqnarray}
where we removed the divergent constant, which appears after the integration
in deriving the second line. 
The divergent constant is a constant phase factor, which may related with the
initial phase, and has no physical meaning.
With the use of the mathematical formula 
\begin{eqnarray}
 e^{iz\sin\theta}=\sum_{s=-\infty}^\infty
e^{is\theta}J_s(z),
\end{eqnarray}
we rewrite Eq.~(\ref{eqgg}) as
\begin{eqnarray}
 g^{\ast}_{p_\rmi}g_p=e^{i\sigma_{\rmi p}k\cdot x/\hbar}
\sum_{\ell=-\infty}^{\infty}e^{i\ell k\cdot x}
C_\ell\left(\xi_{\rmi p}/\hbar,\eta_{\rmi p}/\hbar,
\alpha_{\rmi p},\beta_{\rmi p}\right),
\end{eqnarray}
where we defined 
\begin{eqnarray}
C_\ell\left(\xi_{\rmi p}/\hbar,\eta_{\rmi p}/\hbar,\alpha_{\rmi p}
\beta_{\rmi p}\right)=
\sum_{s=-\infty}^{\infty}e^{i(\ell-2s)\alpha_{\rmi p}}
e^{is\beta_{\rmi p}}J_{\ell-2s}\left(\xi_{\rmi p}/\hbar\right)
J_{s}\left(\eta_{\rmi p}/2\hbar\right).
\end{eqnarray}
%
%
We can write 
\begin{eqnarray}
&& (I^\mu_{p_\rmi}+I^\mu_p+2e\bar{A}^\mu)_{x_1}
(I_{p_\rmi\mu}+I_{p\mu}+2e\bar{A}_\mu)_{x_2}
=\sum_{j,j^\prime=-2}^{2}M_{j,j^\prime}e^{ijk\cdot x_1}
e^{ij^\prime k\cdot x_2},
\end{eqnarray}
where we define the kernel $M_{j,j^\prime}$ as follows,
\begin{eqnarray}
&& M_{0,0}=-2m^2+2p_\rmi\cdot p-\frac{e^2a\cdot a^\ast}{2}(p_\rmi+p)\cdot k
\left(\frac{1}{p_\rmi\cdot k}+\frac{1}{p\cdot k}\right), 
\nonumber \\
&&M_{0,1}=M_{1,0}=M^\ast_{0,-1}=M^\ast_{-1,0}=\frac{\xi_{\rmi p}}{2}
e^{i\alpha_{\rmi p}}(p_\rmi -p)\cdot k,
\nonumber \\
&&M_{1,1}=M^\ast_{-1,-1}=e^2a\cdot a,
~~~~~~
M_{-1,1}=M^\ast_{1,-1}=e^2a\cdot a^\ast
\nonumber \\
&&M_{0,2}=M_{2,0}=M^\ast_{0,-2}=M^\ast_{-2,0}
=-\frac{e^2}{8}(p_\rmi +p)\cdot k
\left(\frac{1}{p_\rmi\cdot k}+\frac{1}{p\cdot k}\right)a\cdot a,
\nonumber
\end{eqnarray}
and the other components of $M_{j,j^\prime}$ are zero. 
By performing the integration with respect to ${\bf x}_2$ 
in Eq.~(\ref{expect}), 
we have the momentum conservation
\begin{eqnarray}
 {\bf p}={\bf p}_\rmi-\hbar{\bf k}_f-{\bf k}[\sigma_{\rmi p}+(\ell+j)\hbar], 
\label{ppkk}
\end{eqnarray}
and, after summing over ${\bf p}$, Eq.~(\ref{expect}) reduces to
\begin{eqnarray}\label{number_momentum}
\sum_{\lambda} \left\langle a^{\dagger\lambda}_{k_f}a^\lambda_{k_f}
 \right\rangle &=& {\rm Re}\Bigl[\frac{\mu_0e^2}{8V\hbar}
\int^{\infty}_{-\infty}dt_2\int^{\infty}_{-\infty}dt_1
\nonumber
\\
&&\sum_{\ell,\ell^\prime,j,j^\prime}
\frac{1}{k_fp^0_ip^0}M_{j,j^\prime}
C_\ell(\xi_{\rmi p}/\hbar,\eta_{\rmi p}/\hbar,\beta_{\rmi p})
C_{\ell^\prime}(\xi_{p\rmi}/\hbar,\eta_{p\rmi}/\hbar,\beta_{\rmi p})\nonumber \\
&\times&
\frac{1}{V}\int d^3{\bf x}_1e^{-i\hbar {\bf k}(\ell+j+\ell^\prime+j^\prime)
\cdot {\bf x}_1/\hbar}
\nonumber \\
&\times&e^{i(p_{0\rmi}-p_0-\hbar k_f-
k[\sigma_{\rmi p}+(\ell+j)\hbar])\cdot t_1/\hbar}
e^{-i(p_{0\rmi}-p_0-\hbar k_f-k[\sigma_{\rmi p}-(\ell^\prime+j^\prime)\hbar])
\cdot t_2/\hbar}\Bigr].
\nonumber
\\\end{eqnarray}
The integration of the second line in Eq.~(\ref{number_momentum}) 
gives ${\bf k}=0$ if $\ell+j+\ell^\prime+j^\prime\neq0$. 
To avoid this, $\ell+j+\ell^\prime+j^\prime=0$ is required 
and the integration gives unity.
Then, summing over $\ell^\prime$ and using the time variable
$T\equiv(t_1+t_2)/2$ and $\Delta t\equiv t_2-t_1$, we obtain
\begin{eqnarray}
 \sum_{\lambda} \left\langle a^{\dagger\lambda}_{k_f}a^\lambda_{k_f}
 \right\rangle &=& {\rm Re}\Bigl[\frac{\mu_0e^2}{8V\hbar}
 \int^{\infty}_{-\infty}dT\int^{\infty}_{-\infty}d\Delta t
\nonumber
\\&&~~~~ \sum_{\ell,j,j^\prime}
 \frac{1}{k_fp^0_ip^0}M_{j,j^\prime}
 C_\ell(\xi_{\rmi p}/\hbar,\eta_{\rmi p}/\hbar,\beta_{\rmi p})
 C_{-\ell-j-j^\prime}(\xi_{p\rmi}/\hbar,\eta_{p\rmi}/\hbar,\beta_{\rmi p})
\nonumber 
\\
&&~~~~
\times e^{-i(p_\rmi-p-\hbar k_f-k[\sigma_{\rmi p}+(\ell+j)\hbar])\cdot
 \Delta t/\hbar}
\Bigr].
\end{eqnarray}
Integration with respect to $\Delta t$ yields the Dirac's delta function
$2\pi\hbar\delta(p_{0\rmi}-p_0-\hbar k_f-k[\sigma_{\rmi p}+(\ell+j)\hbar])$.

The total radiation rate per unit time is given by 
\begin{eqnarray}
&&\frac{dE}{dT}=\frac{d}{dT}\sum_{k_f,\lambda}\hbar k_f
 \left\langle a^{\dagger\lambda}_{k_f}a^\lambda_{k_f}
 \right\rangle
\nonumber
\\
&&~~~~~=
{\rm Re}\Bigl[\frac{\mu_0\hbar e^2}{4(2\pi)^2}\int dk_fd^2\Omega_{k_f}
\frac{k^2_f}{p^0_i}\sum_{\ell,j,j^\prime}M_{j,j^\prime}
C_\ell(\xi_{\rmi p}/\hbar,\eta_{\rmi p}/\hbar,\beta_{\rmi p})
\nonumber\\
&&~~~~~~~~~~~~~~
C_{-\ell-j-j^\prime}(\xi_{p\rmi}/\hbar,\eta_{p\rmi}/\hbar,\beta_{\rmi
p})
\delta(p^2+m^2)\theta(p^0)
\Bigl],
\label{dEdT}
\end{eqnarray}
where 
$d^2\Omega_{k_f}$ is the solid angle in the Fourier space of the wavenumber of emitted photon $k_f$.
In deriving Eq.~(\ref{dEdT}), 
we used the relation
\begin{eqnarray}
 \frac{1}{2p^0}=\int^\infty_{-\infty}dp^0\delta(p^2+m^2)\theta(p^0),
\end{eqnarray}
and performed the integration with respect to $p^0$.
In expression (\ref{dEdT}), $p_0$ should be understood $p_0=p_{0\rmi}-\hbar k_f
-k[\sigma_{\rmi p}+(\ell+j)\hbar]$. Combined with the momentum conservation
(\ref{ppkk}), we have the relation of the energy momentum conservation
\begin{eqnarray}
 p_\mu=p_{\rmi\mu}-\hbar k_{f \mu }-k_\mu[\sigma_{\rmi p}+\hbar(\ell+j)].
\label{ppkk2}
\end{eqnarray}

In the present paper, we consider the following three cases of the 
initial state of the charged particle. 

{\it Case}~(A)~First is the case when the initial momentum of the incident 
charged particle is specified by $p^{\mu}_{\rmi}=(m,0,0,0)$. 
We choose the spatial coordinate $k^\mu=k(1,0,0,1)$ and 
$k_f^\mu=k_f(1,\sin\theta\cos\varphi,
\sin\theta\sin\varphi,\cos\theta)$ for all three cases.  In this case, 
we have 
\begin{eqnarray}
 p_\rmi\cdot k=-mk,\quad p_\rmi\cdot k_f=-mk_f,\quad
k_f\cdot k=-2k_fk\sin^2\frac{\theta}{2}. 
\end{eqnarray}

{\it Case}~(B)~Second is the case when the incident charged particle is
in a relativistic motion, and we choose $p^{\mu}_{\rmi}=(np_i,p_i,0,0)$, 
where we defined $n=\sqrt{1+m^2/p_i^2}$. In this case, 
we have
\begin{eqnarray}
 p_\rmi\cdot k=-np_{\rmi}k,\quad 
p_\rmi\cdot k_f=-p_{\rmi}k_f(n-\sin\theta\cos\phi),\quad
k_f\cdot k=-2k_fk\sin^2\frac{\theta}{2}.
\end{eqnarray}

{\it Case}~(C)~Third is the case when the mean physical momentum of the charged 
particle is zero. Because the mean physical momentum is given by
$q_\rmi^\mu=p^\mu_\rmi-({e^2a\cdot a^\ast}/{4q_\rmi\cdot k})k^\mu$~\footnote{
This expression of the mean momentum is read from the
energy momentum conservation (\ref{ppkk2}), as described in section 4.}, 
then we choose $q^\mu_\rmi=(m,0,0,0)$.
In this case, we have
\begin{eqnarray}
 p_\rmi\cdot k=-mk,\quad 
p_\rmi\cdot k_f=-mk_f+\frac{e^2a\cdot a^\ast}{2m}k_f\sin^2\frac{\theta}{2},\quad
k_f\cdot k=-2k_fk\sin^2\frac{\theta}{2}. 
\end{eqnarray}
In the present paper, we consider the linear polarization for the 
electromagnetic wave background, and assume the constant vector $a^\mu=(0,a,0,0)$. 
Thus, we assume that the motion of the charged particle is  the same 
direction of the polarization vector in case (B).
Due to a fixed electromagnetic wave background, our system is not Lorentz 
invariant. Then, we treat three cases (A), (B), and (C) separately.

We consider the consequence of the relation (\ref{ppkk2}), from which 
we have
\begin{eqnarray}
-(p^2+m^2)&=&2\hbar p_\rmi\cdot k_f+
2\hbar(\ell+j)(p_\rmi\cdot k-\hbar k_f\cdot k)
+2\sigma_{\rmi p}(p_\rmi\cdot k-\hbar k_f\cdot k)
\nonumber \\
&=&2\hbar p_\rmi\cdot k_f+
2\hbar(\ell+j)(p_\rmi\cdot k-\hbar k_f\cdot k)
-\frac{\hbar e^2k_f\cdot k}{2p_\rmi\cdot k}a\cdot a^\ast. 
\label{onshell}
\end{eqnarray}
Because ${p_\rmi}_\mu$ is a time-like vector, then 
$p_\rmi\cdot k_f$ is negative.
Similarly, because $p_\mu$ is time-like, then
$p\cdot k$ is negative, which leads 
$p_i\cdot k -\hbar k_f\cdot k$ must be negative, using Eq.~(\ref{ppkk2}).
Because $k_\mu$ and ${k_{f}}_\mu$ are null vectors, $k\cdot k_f$ is
negative, then $\sigma_{ip}$ is positive. Hence, from $p^2+m^2=0$ with Eq.~(\ref{onshell}), 
$\ell+j$ must be negative. Introducing the positive integer $r=-(\ell+j)\geq 1$,
Eq.~(\ref{ppkk2}) is now written as 
$ p_\mu=p_{\rmi\mu}-\hbar k_{f \mu }-k_\mu(\sigma_{\rmi p}-\hbar r)$.

From the relation $p^2+m^2=0$ with (\ref{onshell}), we have
\begin{eqnarray}
k_f=
\left\{
\begin{array}{ll}
\displaystyle{
\frac{rk}{1+\left(\nu^2+2r\frac{\hbar k}{m}\right)
\sin^2\frac{\theta}{2}}}
&
{\rm ~~~~~{\rm for} ~(A)},\\ 
\displaystyle{
\frac{nrk}{(n-\sin\theta\cos\phi)+
\left(\frac{m^2}{np^2_{\rmi}}\nu^2
+2r\frac{\hbar k}{m}\frac{m}{p_{\rmi}}\right)
\sin^2\frac{\theta}{2}}}
&
{\rm ~~~~~{\rm for} ~(B)},\\ 
\displaystyle{ \frac{rk}{1+2r\frac{\hbar k}{m}
\sin^2\frac{\theta}{2}}}
&
{\rm ~~~~~{\rm for} ~(C)}, 
\end{array}
\right.
\label{kfexp}
\end{eqnarray}
respectively, where we define the dimensionless parameter
\begin{eqnarray}
\nu^2=\frac{e^2}{2m^2}a\cdot a^\ast.
\end{eqnarray}

We may rewrite the expression in Eq.(\ref{dEdT}) as
\begin{eqnarray}
 &&\sum_{\ell=-\infty}^{-j}
C_\ell(\xi_{\rmi p}/\hbar,\eta_{\rmi p}/\hbar,
\alpha_{\rmi p},\beta_{\rmi p})
C_{-\ell-j-j^\prime}(\xi_{p\rmi}/\hbar,\eta_{p\rmi}/\hbar,
\alpha_{\rmi p},\beta_{\rmi
p})\nonumber\\
&&~~~~~~~~~~~~= \sum_{r=1}^{\infty}
C_{-r-j}(\xi_{\rmi p}/\hbar,\eta_{\rmi p}/\hbar,
\alpha_{\rmi p},\beta_{\rmi p})
C_{r-j^\prime}(\xi_{p\rmi}/\hbar,\eta_{p\rmi}/\hbar,
\alpha_{\rmi p},\beta_{\rmi p}),
\label{ccrj}
\end{eqnarray}
then, by performing the integration with respect to $k_f$, we obtain
\begin{eqnarray}\label{energyrate-nonrela}
\frac{d^3E}{d\Omega_{k_f}dT}=
\left\{
\begin{array}{l}
\displaystyle{
\frac{\mu_0 e^2}{8m^2(2\pi)^2}
\sum_{j,j^\prime=-2}^2\sum_{r=1}^\infty
\frac{r^2k^2}{(1+(\nu^2+2r\frac{\hbar k}{m})
\sin^2\frac{\theta}{2})^3}}  \\
\times M_{j,j^\prime}
C_{-r-j}(\xi_{\rmi p}/\hbar,\eta_{\rmi p}/\hbar,
\alpha_{\rmi p},\beta_{\rmi p})
C_{r-j^\prime}(\xi_{p\rmi}/\hbar,\eta_{p\rmi}/\hbar,
\alpha_{\rmi p},\beta_{\rmi
p})
\\
\hspace{8cm}~~~~~{\rm {\rm for}~(A)},\\
\displaystyle{
\frac{\mu_0 e^2}{8m^2(2\pi)^2}
\sum_{j,j^\prime=-2}^2\sum_{r=1}^\infty
\frac{n^2r^2k^2}{(n-\sin\theta\cos\phi+
(\frac{m^2}{np_\rmi^2}\nu^2+2r\frac{\hbar k}{m}\frac{m}{p_\rmi})
\sin^2\frac{\theta}{2})^3}}  \\
\times M_{j,j^\prime}
C_{-r-j}(\xi_{\rmi p}/\hbar,\eta_{\rmi p}/\hbar,
\alpha_{\rmi p},\beta_{\rmi p})
C_{r-j^\prime}(\xi_{p\rmi}/\hbar,\eta_{p\rmi}/\hbar,
\alpha_{\rmi p},\beta_{\rmi
p})
\\ 
\hspace{8cm}~~~~~{\rm {\rm for}~(B)},\\
\displaystyle{
\frac{\mu_0 e^2}{8m^2(2\pi)^2}
\sum_{j,j^\prime=-2}^2\sum_{r=1}^\infty
\frac{r^2k^2}{(1+2r\frac{\hbar k}{m}
\sin^2\frac{\theta}{2})^3}} \\
\times M_{j,j^\prime}
C_{-r-j}(\xi_{\rmi p}/\hbar,\eta_{\rmi p}/\hbar,
\alpha_{\rmi p},\beta_{\rmi p})
C_{r-j^\prime}(\xi_{p\rmi}/\hbar,\eta_{p\rmi}/\hbar,
\alpha_{\rmi p},\beta_{\rmi
p})
\\
\hspace{8cm}~~~~~{\rm {\rm for}~(C)},
\end{array}
\right.
\nonumber\\
\end{eqnarray}
respectively, where $k_f$ should be regarded as the expression 
of the right-hand-side of (\ref{kfexp}). Here we omit the step 
function $\theta(p^0)$ because of the time-like property of $p^\mu$.
When the background electromagnetic wave is linearly polarized 
and the constant vector is $a^\mu=(0,a,0,0)$,  
we have 
\begin{eqnarray}
\displaystyle{
 {\sigma_{\rmi p}\over \hbar}=
 {\eta_{\rmi p}\over \hbar}=
 -{\eta_{p\rmi}\over \hbar}=}
\left\{
\begin{array}{ll}
\displaystyle{
\frac{r\nu^2\sin^2\frac{\theta}{2}}{1+\nu^2\sin^2\frac{\theta}{2}}} &
~~~~~~~{\rm {\rm for}~(A)},
\\
\displaystyle{
\frac{r\nu^2\sin^2\frac{\theta}{2}}
{n\frac{p^2_{\rmi}}{m^2}(n-\sin\theta\cos\phi)
+\nu^2\sin^2\frac{\theta}{2}}} &
~~~~~~~{\rm {\rm for}~(B)},
\\
\displaystyle{
r \nu^2  \sin^2\frac{\theta}{2}} &
~~~~~~~{\rm {\rm for}~(C)},
\end{array}
\right.
\end{eqnarray}
and 
\begin{eqnarray}
 \displaystyle{
 {\xi_{\rmi p}\over \hbar}=
 -{\xi_{p\rmi}\over \hbar}=}
\left\{
\begin{array}{ll}
\displaystyle{
-\frac{\sqrt{2}r\nu\sin\theta\cos\phi}{1+\nu^2\sin^2\frac{\theta}{2}}} &
~~~~~~~~~~~~~~{\rm {\rm for}~(A)},\\
 \displaystyle{
-\frac{\sqrt{2}r\frac{m}{p_\rmi}\nu n
\left(2\sin^2\frac{\theta}{2}-\sin\theta\cos\phi\right)}
{n(n-\sin\theta\cos\phi)
+\frac{m^2}{p^2_{\rmi}}\nu^2\sin^2\frac{\theta}{2}}} &
~~~~~~~~~~~~~~{\rm {\rm for}~(B)},\\
-\sqrt{2}r\nu \sin\theta\cos\phi &
~~~~~~~~~~~~~~{\rm {\rm for}~(C)},\\
\end{array}
\right.
\end{eqnarray}
respectively. It is worthy to note 
that the expression (\ref{ccrj}) does not contain $\hbar$. 

\section{results}
In this section, we present the explicit formulas for the radiation
in three cases, then discuss about the origin of the quantum effect 
by extracting the term of the order of $\hbar$.
\subsection{$p_i^\mu =(m,0,0,0)$}
First let us consider {\it case} (A). Explicit expression of
the kernel $M_{i,j}$ is summarized in Appendix A. 
Expression (\ref{energyrate-nonrela}) yields
\begin{eqnarray}\label{rad-nonrela}
&& \frac{d^3E}{d\Omega_{k_f}dT}=
\frac{\mu_0 e^2}{8(2\pi)^2}
\sum^\infty_{s,s^\prime=-\infty}\sum_{r=1}^\infty
\frac{r^2k^2}{(1+(\nu^2+2r\frac{\hbar k}{m})
\sin^2\frac{\theta}{2})^3}\nonumber \\
&\times&\Bigl\{
-4\frac{1+\nu^2+
\left(\nu^4+\frac{\hbar kr}{m}(2+\frac{\hbar kr}{m})
+\nu^2(1+2\frac{\hbar kr}{m})\right)\sin^2\frac{\theta}{2}}
{1+(\nu^2+2\frac{\hbar kr}{m})\sin^2\frac{\theta}{2}}
\nonumber \\
&\times&
J_{-r-2s}(\xi_{\rmi p}/\hbar)J_{s}(\eta_{\rmi p}/2\hbar)
J_{r-2s^\prime}(\xi_{p\rmi}/\hbar)J_{s^\prime}(\eta_{p\rmi}/2\hbar)
\nonumber \\
&+&\frac{\sqrt{2}\nu\frac{\hbar^2 k^2r^2}{m^2}\cos\phi\sin^2\frac{\theta}{2}
\sin\theta}{\left(1+\nu^2\sin^2\frac{\theta}{2}\right)
\left(1+\left(\nu^2+2\frac{\hbar
	 kr}{m}\right)\sin^2\frac{\theta}{2}\right)}
\nonumber \\
&\times&\Bigl(
J_{-r-1-2s}(\xi_{\rmi p}/\hbar)J_{s}(\eta_{\rmi p}/2\hbar)
J_{r-2s^\prime}(\xi_{p\rmi}/\hbar)J_{s^\prime}(\eta_{p\rmi}/2\hbar)
\nonumber \\
&+&
J_{-r+1-2s}(\xi_{\rmi p}/\hbar)J_{s}(\eta_{\rmi p}/2\hbar)
J_{r-2s^\prime}(\xi_{p\rmi}/\hbar)J_{s^\prime}(\eta_{p\rmi}/2\hbar)
\nonumber \\
&+&
J_{-r-2s}(\xi_{\rmi p}/\hbar)J_{s}(\eta_{\rmi p}/2\hbar)
J_{r-1-2s^\prime}(\xi_{p\rmi}/\hbar)J_{s^\prime}(\eta_{p\rmi}/2\hbar)
\nonumber \\
&+&
J_{-r-2s}(\xi_{\rmi p}/\hbar)J_{s}(\eta_{\rmi p}/2\hbar)
J_{r+1-2s^\prime}(\xi_{p\rmi}/\hbar)J_{s^\prime}(\eta_{p\rmi}/2\hbar)
\Bigr)
\nonumber \\
&+&2\nu^2\Bigl(
J_{-r-1-2s}(\xi_{\rmi p}/\hbar)J_{s}(\eta_{\rmi p}/2\hbar)
J_{r-1-2s^\prime}(\xi_{p\rmi}/\hbar)J_{s^\prime}(\eta_{p\rmi}/2\hbar)
\nonumber \\
&+&
J_{-r+1-2s}(\xi_{\rmi p}/\hbar)J_{s}(\eta_{\rmi p}/2\hbar)
J_{r+1-2s^\prime}(\xi_{p\rmi}/\hbar)J_{s^\prime}(\eta_{p\rmi}/2\hbar)
\nonumber \\
&+&
J_{-r-1-2s}(\xi_{\rmi p}/\hbar)J_{s}(\eta_{\rmi p}/2\hbar)
J_{r+1-2s^\prime}(\xi_{p\rmi}/\hbar)J_{s^\prime}(\eta_{p\rmi}/2\hbar)
\nonumber \\
&+&
J_{-r+1-2s}(\xi_{\rmi p}/\hbar)J_{s}(\eta_{\rmi p}/2\hbar)
J_{r-1-2s^\prime}(\xi_{p\rmi}/\hbar)J_{s^\prime}(\eta_{p\rmi}/2\hbar)
\Bigr)
\nonumber \\
&-&
\frac{\nu^2\left(1+\left(\nu^2+\frac{\hbar kr}{m}\right)
\sin^2\frac{\theta}{2}\right)^2}{\left(1+\nu^2\sin^2\frac{\theta}{2}\right)
\left(1+\left(\nu^2+2\frac{\hbar kr }{m}\right)\sin^2\frac{\theta}{2}\right)}
\nonumber \\
&\times&
\Bigl(
J_{-r-2-2s}(\xi_{\rmi p}/\hbar)J_{s}(\eta_{\rmi p}/2\hbar)
J_{r-2s^\prime}(\xi_{p\rmi}/\hbar)J_{s^\prime}(\eta_{p\rmi}/2\hbar)
\nonumber \\
&+&
J_{-r+2-2s}(\xi_{\rmi p}/\hbar)J_{s}(\eta_{\rmi p}/2\hbar)
J_{r-2s^\prime}(\xi_{p\rmi}/\hbar)J_{s^\prime}(\eta_{p\rmi}/2\hbar)
\nonumber \\
&+&
J_{-r-2s}(\xi_{\rmi p}/\hbar)J_{s}(\eta_{\rmi p}/2\hbar)
J_{r-2-2s^\prime}(\xi_{p\rmi}/\hbar)J_{s^\prime}(\eta_{p\rmi}/2\hbar)
\nonumber \\
&+&
J_{-r-2s}(\xi_{\rmi p}/\hbar)J_{s}(\eta_{\rmi p}/2\hbar)
J_{r+2-2s^\prime}(\xi_{p\rmi}/\hbar)J_{s^\prime}(\eta_{p\rmi}/2\hbar)
\Bigr)\Bigr\}
.
\end{eqnarray}
When the background field is weak, i.e., $\nu\ll 1$, one can verify
that $r=1$ gives the dominant contribution. In this case, we find the first 
order quantum effect by expanding 
the above expression up to $O(\nu^2)$ and $O(\hbar)$,
\begin{eqnarray}
 && \frac{d^3E}{d\Omega_{k_f}dT}=
\frac{\mu_0 e^2k^2\nu^2}{(4\pi)^2}
(1-6\frac{\hbar k}{m}
\sin^2\frac{\theta}{2})
\left(1-\cos^2\phi\sin^2\theta\right)
.
\end{eqnarray}
The integration with respect to $\Omega_{k_f}$ gives
\begin{eqnarray}
\frac{dE}{dT}=
\frac{\mu_0 e^2k^2\nu^2}{6\pi}(1-3\frac{\hbar k}{m}),\label{firstnonrela}
\end{eqnarray}
which means that the first-order quantum effect decreases the radiation energy.

\subsection{$p_i^\mu =(np_i,p_i,0,0)$}
In {\it case} (B), expression (\ref{energyrate-nonrela}) leads to
\begin{eqnarray}\label{rad-rela}
&& \frac{d^3E}{d\Omega_{k_f}dT}=
\frac{\mu_0 e^2}{8(2\pi)^2}
\sum^\infty_{s,s^\prime=-\infty}\sum_{r=1}^\infty
\frac{n^2r^2k^2}{\left((n-\sin\theta\cos\phi)+(\frac{m^2}{np^2_{\rmi}}\nu^2
+2r\frac{\hbar k}{m}\frac{m}{p_\rmi})
\sin^2\frac{\theta}{2}\right)^3}\nonumber \\
&\times&\Bigl\{
-4\frac{n(1+\nu^2)(n-\sin\theta\cos\phi)+
\left(\frac{m^2}{p^2_{\rmi}}(\nu^2+\nu^4)+2n\frac{\hbar kr}{m}
\frac{m}{p_{\rmi}}(1+\nu^2)+
n^2\frac{\hbar^2k^2r^2}{m^2}\right)\sin^2\frac{\theta}{2}}
{n(n-\sin\theta\cos\phi)+\left(\frac{m^2}{p^2_{\rmi}}\nu^2+
2n\frac{\hbar kr}{m}\frac{m}{p_{\rmi}}\right)\sin^2\frac{\theta}{2}}
\nonumber \\
&\times&
J_{-r-2s}(\xi_{\rmi p}/\hbar)J_{s}(\eta_{\rmi p}/2\hbar)
J_{r-2s^\prime}(\xi_{p\rmi}/\hbar)J_{s^\prime}(\eta_{p\rmi}/2\hbar)
\nonumber\\
&+&
\sqrt{2}
\frac{n^3\nu\frac{\hbar^2k^2r^2}{m^2}\frac{m}{p_{\rmi}}\sin^2\frac{\theta}{2}
\left(2\sin^2\frac{\theta}{2}-\sin\theta\cos\phi\right)}
{\left(n(n-\sin\theta\cos\phi)+
\left(\frac{m^2}{p^2_{\rmi}}\nu^2+2n\frac{\hbar
 kr}{m}\frac{m}{p_{\rmi}}\right)\sin^2\frac{\theta}{2}\right)
\left(n(n-\sin\theta\cos\phi)+\frac{m^2}{p^2_{\rmi}}\nu^2\sin^2\frac{\theta}{2}\right)}
\nonumber \\
&\times&\Bigl(
J_{-r-1-2s}(\xi_{\rmi p}/\hbar)J_{s}(\eta_{\rmi p}/2\hbar)
J_{r-2s^\prime}(\xi_{p\rmi}/\hbar)J_{s^\prime}(\eta_{p\rmi}/2\hbar)
\nonumber \\
&+&
J_{-r+1-2s}(\xi_{\rmi p}/\hbar)J_{s}(\eta_{\rmi p}/2\hbar)
J_{r-2s^\prime}(\xi_{p\rmi}/\hbar)J_{s^\prime}(\eta_{p\rmi}/2\hbar)
\nonumber \\
&+&
J_{-r-2s}(\xi_{\rmi p}/\hbar)J_{s}(\eta_{\rmi p}/2\hbar)
J_{r-1-2s^\prime}(\xi_{p\rmi}/\hbar)J_{s^\prime}(\eta_{p\rmi}/2\hbar)
\nonumber \\
&+&
J_{-r-2s}(\xi_{\rmi p}/\hbar)J_{s}(\eta_{\rmi p}/2\hbar)
J_{r+1-2s^\prime}(\xi_{p\rmi}/\hbar)J_{s^\prime}(\eta_{p\rmi}/2\hbar)
\Bigr)
\nonumber \\
&+&2\nu^2\Bigl(
J_{-r-1-2s}(\xi_{\rmi p}/\hbar)J_{s}(\eta_{\rmi p}/2\hbar)
J_{r-1-2s^\prime}(\xi_{p\rmi}/\hbar)J_{s^\prime}(\eta_{p\rmi}/2\hbar)
\nonumber \\
&+&
J_{-r+1-2s}(\xi_{\rmi p}/\hbar)J_{s}(\eta_{\rmi p}/2\hbar)
J_{r+1-2s^\prime}(\xi_{p\rmi}/\hbar)J_{s^\prime}(\eta_{p\rmi}/2\hbar)
\nonumber \\
&+&
J_{-r-1-2s}(\xi_{\rmi p}/\hbar)J_{s}(\eta_{\rmi p}/2\hbar)
J_{r+1-2s^\prime}(\xi_{p\rmi}/\hbar)J_{s^\prime}(\eta_{p\rmi}/2\hbar)
\nonumber \\
&+&
J_{-r+1-2s}(\xi_{\rmi p}/\hbar)J_{s}(\eta_{\rmi p}/2\hbar)
J_{r-1-2s^\prime}(\xi_{p\rmi}/\hbar)J_{s^\prime}(\eta_{p\rmi}/2\hbar)
\Bigr)
\nonumber \\
&-&
\frac{\nu^2}{4}
\frac{\left(2n(n-\sin\theta\cos\phi)+\frac{m^2}{p^2_{\rmi}}\nu^2+
n\frac{\hbar kr}{m}\frac{m}{p_{\rmi}}-
\left(\frac{m^2}{p^2_{\rmi}}\nu^2+n\frac{\hbar
 kr}{m}\frac{m}{p_{\rmi}}\right)\cos\theta
\right)^2}
{\left(n(n-\sin\theta\cos\phi)+\frac{m^2}{p^2_{\rmi}}
\nu^2\sin^2\frac{\theta}{2}\right)
\left(n(n-\sin\theta\cos\phi)+
\left(\frac{m^2}{p^2_{\rmi}}\nu^2+2n\frac{\hbar kr}{m}\frac{m}{p_{\rmi}}
\right)\sin^2\frac{\theta}{2}\right)}
\nonumber \\
&\times&
\Bigl(
J_{-r-2-2s}(\xi_{\rmi p}/\hbar)J_{s}(\eta_{\rmi p}/2\hbar)
J_{r-2s^\prime}(\xi_{p\rmi}/\hbar)J_{s^\prime}(\eta_{p\rmi}/2\hbar)
\nonumber \\
&+&
J_{-r+2-2s}(\xi_{\rmi p}/\hbar)J_{s}(\eta_{\rmi p}/2\hbar)
J_{r-2s^\prime}(\xi_{p\rmi}/\hbar)J_{s^\prime}(\eta_{p\rmi}/2\hbar)
\nonumber \\
&+&
J_{-r-2s}(\xi_{\rmi p}/\hbar)J_{s}(\eta_{\rmi p}/2\hbar)
J_{r-2-2s^\prime}(\xi_{p\rmi}/\hbar)J_{s^\prime}(\eta_{p\rmi}/2\hbar)
\nonumber \\
&+&
J_{-r-2s}(\xi_{\rmi p}/\hbar)J_{s}(\eta_{\rmi p}/2\hbar)
J_{r+2-2s^\prime}(\xi_{p\rmi}/\hbar)J_{s^\prime}(\eta_{p\rmi}/2\hbar)
\Bigr)\Bigr\}
.
\end{eqnarray}
Expanding this expression up to $O(\nu^2)$, $O(\hbar)$ and
$O({m}/{p_\rmi})$ in the relativistic limit, we obtain
\begin{eqnarray}
  \frac{d^3E}{d\Omega_{k_f}dT}=
\frac{\mu_0 k^2\nu^2e^2}{4(2\pi)^2}
\left(\frac{1}{(1-\sin\theta\cos\phi)^3}-6\frac{\hbar k}{m}
\frac{m}{p_\rmi}\frac{\sin^2\frac{\theta}{2}}{(1-\sin\theta\cos\phi)^4}\right).
\end{eqnarray}
The integration with respect to $\Omega_{k_f}$ gives
\begin{eqnarray}
  \frac{dE}{dT}=
\frac{\mu_0 k^2e^2\nu^2}{32\pi}
\left(3-7\frac{\hbar k}{m}\frac{m}{p_\rmi}\right).\label{firstrela}
\end{eqnarray}

\subsection{$q_i^\mu=(m,0,0,0)$}
In {\it case} (C), expression (\ref{energyrate-nonrela}) yields
\begin{eqnarray}\label{rad-rest}
&& \frac{d^3E}{d\Omega_{k_f}dT}=
\frac{\mu_0 e^2}{8(2\pi)^2}
\sum^\infty_{s,s^\prime=-\infty}\sum_{r=1}^\infty
\frac{r^2k^2}{(1+2r\frac{\hbar k}{m}
\sin^2\frac{\theta}{2})^3}\nonumber \\
&\times&\Bigl\{
-4\frac{1+\nu^2+\frac{\hbar kr}{m}
(2(1+\nu^2)+\frac{\hbar kr}{m})\sin^2\frac{\theta}{2}}
{1+2\frac{\hbar kr}{m}\sin^2\frac{\theta}{2}}
\nonumber \\
&\times&
J_{-r-2s}(\xi_{\rmi p}/\hbar)J_{s}(\eta_{\rmi p}/2\hbar)
J_{r-2s^\prime}(\xi_{p\rmi}/\hbar)J_{s^\prime}(\eta_{p\rmi}/2\hbar)
+\frac{\sqrt{2}\nu\frac{\hbar^2 k^2r^2}{m^2}\cos\phi\sin^2\frac{\theta}{2}
\sin\theta}{1+2\frac{\hbar kr}{m}\sin^2\frac{\theta}{2}}
\nonumber \\
&\times&\Bigl(
J_{-r-1-2s}(\xi_{\rmi p}/\hbar)J_{s}(\eta_{\rmi p}/2\hbar)
J_{r-2s^\prime}(\xi_{p\rmi}/\hbar)J_{s^\prime}(\eta_{p\rmi}/2\hbar)
\nonumber \\
&+&
J_{-r+1-2s}(\xi_{\rmi p}/\hbar)J_{s}(\eta_{\rmi p}/2\hbar)
J_{r-2s^\prime}(\xi_{p\rmi}/\hbar)J_{s^\prime}(\eta_{p\rmi}/2\hbar)
\nonumber \\
&+&
J_{-r-2s}(\xi_{\rmi p}/\hbar)J_{s}(\eta_{\rmi p}/2\hbar)
J_{r-1-2s^\prime}(\xi_{p\rmi}/\hbar)J_{s^\prime}(\eta_{p\rmi}/2\hbar)
\nonumber \\
&+&
J_{-r-2s}(\xi_{\rmi p}/\hbar)J_{s}(\eta_{\rmi p}/2\hbar)
J_{r+1-2s^\prime}(\xi_{p\rmi}/\hbar)J_{s^\prime}(\eta_{p\rmi}/2\hbar)
\Bigr)
\nonumber \\
&+&2\nu^2\Bigl(
J_{-r-1-2s}(\xi_{\rmi p}/\hbar)J_{s}(\eta_{\rmi p}/2\hbar)
J_{r-1-2s^\prime}(\xi_{p\rmi}/\hbar)J_{s^\prime}(\eta_{p\rmi}/2\hbar)
\nonumber \\
&+&
J_{-r+1-2s}(\xi_{\rmi p}/\hbar)J_{s}(\eta_{\rmi p}/2\hbar)
J_{r+1-2s^\prime}(\xi_{p\rmi}/\hbar)J_{s^\prime}(\eta_{p\rmi}/2\hbar)
\nonumber \\
&+&
J_{-r-1-2s}(\xi_{\rmi p}/\hbar)J_{s}(\eta_{\rmi p}/2\hbar)
J_{r+1-2s^\prime}(\xi_{p\rmi}/\hbar)J_{s^\prime}(\eta_{p\rmi}/2\hbar)
\nonumber \\
&+&
J_{-r+1-2s}(\xi_{\rmi p}/\hbar)J_{s}(\eta_{\rmi p}/2\hbar)
J_{r-1-2s^\prime}(\xi_{p\rmi}/\hbar)J_{s^\prime}(\eta_{p\rmi}/2\hbar)
\Bigr)
\nonumber \\
&-&
\frac{\nu^2\left(1+\frac{\hbar kr}{m}\sin^2\frac{\theta}{2}
\right)^2}{1+2\frac{\hbar kr }{m}\sin^2\frac{\theta}{2}}
\Bigl(
J_{-r-2-2s}(\xi_{\rmi p}/\hbar)J_{s}(\eta_{\rmi p}/2\hbar)
J_{r-2s^\prime}(\xi_{p\rmi}/\hbar)J_{s^\prime}(\eta_{p\rmi}/2\hbar)
\nonumber \\
&+&
J_{-r+2-2s}(\xi_{\rmi p}/\hbar)J_{s}(\eta_{\rmi p}/2\hbar)
J_{r-2s^\prime}(\xi_{p\rmi}/\hbar)J_{s^\prime}(\eta_{p\rmi}/2\hbar)
\nonumber \\
&+&
J_{-r-2s}(\xi_{\rmi p}/\hbar)J_{s}(\eta_{\rmi p}/2\hbar)
J_{r-2-2s^\prime}(\xi_{p\rmi}/\hbar)J_{s^\prime}(\eta_{p\rmi}/2\hbar)
\nonumber \\
&+&
J_{-r-2s}(\xi_{\rmi p}/\hbar)J_{s}(\eta_{\rmi p}/2\hbar)
J_{r+2-2s^\prime}(\xi_{p\rmi}/\hbar)J_{s^\prime}(\eta_{p\rmi}/2\hbar)
\Bigr)\Bigr\}
.
\end{eqnarray}
Expanding up to $O(\hbar)$ and $O(\nu^2)$, we find
\begin{eqnarray}
 \frac{d^3E}{d\Omega_{k_f}dT}=
\frac{\mu_0 k^2\nu^2 e^2}{16(2\pi)^2}
(3+2\cos 2\theta\cos^2\phi-\cos2\phi)
(1+6\frac{\hbar k }{m}\sin^2\frac{\theta}{2}).
\end{eqnarray}
The integration with respect to $\Omega_{k_f}$ gives
\begin{eqnarray}
  \frac{dE}{dT}=
\frac{\mu_0 k^2e^2\nu^2}{12\pi}
\left(1-3\frac{\hbar k}{m}\right).
\label{firstrest}
\end{eqnarray}

Let us summarize the results in this section. 
We presented the explicit formulas
for the radiation rate for three cases. 
In all three cases, the first order quantum effect decreases 
the total radiation energy.
Because $\xi_{ip}/\hbar$ and $\eta_{ip}/\hbar$ do not depend 
on $\hbar$, then the Bessel functions in the radiation 
formulas do not contain $\hbar$. 
As demonstrated in appendix A, the quantum correction in 
the kernel $M_{i,j}$ arises at the order of $\hbar^2$.  
Then, the first order correction of $\hbar$ comes from the first 
line of the radiation formula (\ref{rad-nonrela}), 
(\ref{rad-rela}) and (\ref{rad-rest}), respectively.

\section{summary and conclusions}
In the present paper, we investigated the quantum effect on the 
Larmor radiation from a moving charge in a monochromatic 
electromagnetic plane wave based on the SQED.  
Our work is different from the previous works 
in the following points. First, we derived the theoretical formula 
starting with the framework of the in-in formalism\cite{Wein}. 
Second, we demonstrated that the quantum effect generally suppresses the total radiation 
energy by explicitly evaluating the contribution at the order of $\hbar$. 
To this end, we considered the three cases for the initial state,
(A) the initial momentum of the charged particle is zero, 
(B) the charged particle is in the relativistic motion 
along with the direction of the linear polarization of 
electromagnetic plane wave, and (C) the mean momentum is zero.
We derived the formula for the radiation per unit time 
(\ref{rad-nonrela}), (\ref{rad-rela}) and (\ref{rad-rest}), 
for (A), (B) and (C), respectively.
The results contain the two parameters,
the strength of the electromagnetic plane wave $\nu$ and  
the frequency of the electromagnetic plane wave ${\hbar k}/{m}$
for (A) and (C). 
One more parameter, the initial momentum of the charged 
particle ${m}/{p_i}$ is added for (B).
We find that the first order quantum correction is the order of 
${\hbar k}/{m}$ 
for (A) and (C), and is the order 
of ${\hbar k}/{m}$  multiplied by the additional factor $m/p_i$ 
for (B).
Thus, the first order quantum effect is the order ${\hbar k}/{m}$ at most.

The quantum effect decreases the total radiation energy compared 
with the classical radiation formula for three cases. 
Let us discuss about the origin of the quantum effect.
As described in the previous sections, the quantum effect in $M_{ij}$ appears
at the order of $\hbar^2$ and the Bessel function doesn't contain
$\hbar$ in their arguments. The first order quantum correction comes from
the denominator of the factor in the first line of 
Eqs.~(\ref{rad-nonrela}), (\ref{rad-rela}) and (\ref{rad-rest}), 
for (A), (B) and (C), respectively.
This factor comes from the scattered photon energy (\ref{kfexp})
and the energy-momentum conservation (\ref{ppkk2}), 
which is rewritten as,
\begin{eqnarray}
p_\mu-{e^2|a|^2\over 4p\cdot k}k_\mu+\hbar {k_f}_\mu
 ={p_\rmi}_\mu-{e^2|a|^2\over 4p_i\cdot k}k_\mu+\hbar r k_\mu.
\end{eqnarray}
This represents the energy-momentum conservation of the (nonlinear)
Compton scattering as $p_\mu-(e^2|a|^2/4p\cdot k)k_\mu$ is regarded as 
the mean momentum of charged particle averaged over long duration of time. 
Therefore, the first order quantum effect originates {\it only} from 
the inelastic collision of the Compton scattering.
This explains the decrease of the energy of a scattered photon 
and the total radiation energy. 
Thus, the particle properties of a photon is the 
origin of the quantum effect in the case of the present paper. 

Let us discuss the differences between the results in the present 
paper and those in the previous works\cite{NSY,Higuchi,YN,KNY}. 
First, in the previous works, the first order quantum effect 
is identified as the non-local nature of the charged particle, 
which is a sharp contrast to the conclusion of the present work. 
The previous works considered a spatially homogeneous 
electric field background, in which we could not regard the radiation 
process as a collision between a charged particle and background photons.
Second, in our previous work\cite{YN} based on the Wentzel-Kramers-Brillouin (WKB)
approximation, the quantum effect may increase the total 
radiation when the charged particle is in the relativistic 
motion in an oscillating homogeneous electric field.
This difference of the results can be ascribed to the 
difference of the origin of the quantum effect. 

Our theoretical framework, based on the SQED, might be useful as a tool to investigate the 
quantum radiation from an electron in an intense electromagnetic field.
When we assume a charged particle as an electron and 
the electromagnetic plane wave background as an X-ray laser, we have 
\begin{eqnarray}
 \frac{\hbar k}{m}\simeq  
 2\times10^{-3}\left({\hbar k\over 1{\rm eV}}\right)
 \left({0.5 {\rm MeV}\over m}\right).
\end{eqnarray}
Thus, the quantum effect is very small. This will make it hard to 
detect the quantum effect in an experiment.
However, the interaction between the laser and an electron has attracted 
many researchers even in the area of the fundamental physics (see e.g., 
Refs.~\refcite{CT,Iso} and references therein).
Further investigations on the topic, how the quantum effect could be tested 
in experiments, are left as a future problem. 

\vspace{3mm}
{\it Acknowledgment} 
We thank S.~Iso, S.~Zhang, T. Kato, K.~Homma and T.~Takahashi for useful conversation related to the topic 
in the present paper. This work was supported by a Grant-in-Aid for Scientific
research of the Japanese Ministry of Education,
Culture, Sports, Science and Technology (No. 21540270 and No.~21244033),
and in part by the Japan Society for Promotion of Science
(JSPS) Core-to-Core Program ``International Research
Network for Dark Energy.''
G.N. was supported by Grant-in-Aid for Japan Society for
Promotion of Science (JSPS) Fellows (No.~236669).

\begin{appendix}
\section{Expression of kernel}
In the appendix, we summarize the explicit expression of
the kernel $M_{ij}$. Straightforward calculation leads to
\begin{eqnarray}
&& M_{0,0}=-4m^2\frac{1+\nu^2+
\left(\nu^4+\frac{\hbar kr}{m}(2+\frac{\hbar kr}{m})
+\nu^2(1+2\frac{\hbar kr}{m})\right)\sin^2\frac{\theta}{2}}
{1+(\nu^2+2\frac{\hbar kr}{m})\sin^2\frac{\theta}{2}}
\nonumber\\
&&M_{1,0}=M_{0,1}=M_{-1,0}=M_{0,-1}=\sqrt{2}m^2
\frac{\nu\frac{\hbar^2k^2r^2}{m^2}\cos\phi\sin^2
\frac{\theta}{2}\sin\theta}
{\left(1+\nu^2\sin^2\frac{\theta}{2}\right)
\left(1+\left(\nu^2+2\frac{\hbar kr}{m}\right)\sin^2\frac{\theta}{2}\right)}
\nonumber\\
&&M_{1,1}=M_{-1,-1}=M_{-1,1}=M_{1,-1}=2m^2\nu^2
\nonumber\\
&&M_{2,0}=M_{0,2}=M_{-2,0}=M_{0,-2}=
-m^2\nu^2\frac{\left(1+\left(\nu^2+\frac{\hbar kr}{m}\right)
\sin^2\frac{\theta}{2}\right)^2}
{\left(1+\nu^2\sin^2\frac{\theta}{2}\right)
\left(1+\left(\nu^2+2\frac{\hbar kr}{m}\right)\sin^2\frac{\theta}{2}\right)}
\nonumber
\end{eqnarray}
for (A),  
\begin{eqnarray}
&& M_{0,0}=-4m^2\frac{n(1+\nu^2)(n-\sin\theta\cos\phi)+
\left(\frac{m^2}{p^2_{\rmi}}(\nu^2+\nu^4)+2n\frac{\hbar kr}{m}
\frac{m}{p_{\rmi}}(1+\nu^2)+
n^2\frac{\hbar^2k^2r^2}{m^2}\right)\sin^2\frac{\theta}{2}}
{n(n-\sin\theta\cos\phi)+\left(\frac{m^2}{p^2_{\rmi}}\nu^2+
2n\frac{\hbar kr}{m}\frac{m}{p_{\rmi}}\right)\sin^2\frac{\theta}{2}}\nonumber\\
&&M_{1,0}=M_{0,1}=M_{-1,0}=M_{0,-1}\nonumber\\\nonumber\\
&&=\sqrt{2}m^2
\frac{n^3\nu\frac{\hbar^2k^2r^2}{m^2}\frac{m}{p_{\rmi}}\sin^2\frac{\theta}{2}
\left(2\sin^2\frac{\theta}{2}-\sin\theta\cos\phi\right)}
{\left(n(n-\sin\theta\cos\phi)+
\left(\frac{m^2}{p^2_{\rmi}}\nu^2+2n\frac{\hbar
 kr}{m}\frac{m}{p_{\rmi}}\right)\sin^2\frac{\theta}{2}\right)
\left(n(n-\sin\theta\cos\phi)+\frac{m^2}{p^2_{\rmi}}\nu^2\sin^2\frac{\theta}{2}\right)}
\nonumber\\
&&M_{1,1}=M_{-1,-1}=M_{-1,1}=M_{1,-1}=2m^2\nu^2
\nonumber\\
&&M_{2,0}=M_{0,2}=M_{-2,0}=M_{0,-2}\nonumber\\\nonumber\\
&&=
-m^2\nu^2
\frac{\left(n(n-\sin\theta\cos\phi)+
\left(\frac{m^2}{p^2_{\rmi}}\nu^2+n\frac{\hbar
 kr}{m}\frac{m}{p_{\rmi}}\right)\sin^2\frac{\theta}{2}
\right)^2}
{\left(n(n-\sin\theta\cos\phi)+\frac{m^2}{p^2_{\rmi}}
\nu^2\sin^2\frac{\theta}{2}\right)
\left(n(n-\sin\theta\cos\phi)+
\left(\frac{m^2}{p^2_{\rmi}}\nu^2+2n\frac{\hbar kr}{m}\frac{m}{p_{\rmi}}
\right)\sin^2\frac{\theta}{2}\right)}
\nonumber
\end{eqnarray}
for (B), and
\begin{eqnarray}
 && M_{0,0}=-4m^2\frac{1+\nu^2+\frac{\hbar k
  r}{m}\left(2(1+\nu^2)+\frac{\hbar kr}{m}\right)\sin^2\frac{\theta}{2}}
{1+2\frac{\hbar kr}{m}\sin^2\frac{\theta}{2}}\nonumber\\
&&M_{1,0}=M_{0,1}=M_{-1,0}=M_{0,-1}
=\sqrt{2}m^2\nu\frac{\frac{\hbar^2k^2r^2}{m^2}
\sin\theta\cos\phi\sin^2\frac{\theta}{2}}
{1+2\frac{\hbar k r}{m}\sin^2\frac{\theta}{2}}
\nonumber\\
&&M_{1,1}=M_{-1,-1}=M_{-1,1}=M_{1,-1}=2m^2\nu^2
\nonumber\\
&&M_{2,0}=M_{0,2}=M_{-2,0}=M_{0,-2}=
-m^2\frac{\nu^2\left(1+\frac{\hbar k
		r}{m}\sin^2\frac{\theta}{2}\right)^2}
{1+2\frac{\hbar kr}{m}\sin^2\frac{\theta}{2}}
\nonumber
\end{eqnarray}
for (C), respectively. It is worthy to note the 
expressions of $M_{i,j}$ expanded in terms of $\hbar$. 
From the above formulas, we have
\begin{eqnarray}
  && M_{0,0}=-4m^2(1+\nu^2)-4m^2\frac{\hbar^2 k^2 r^2}{m^2}
\frac{\sin^2\frac{\theta}{2}}{1+\nu^2\sin^2\frac{\theta}{2}}+O(\hbar^3)\nonumber\\
&&M_{1,0}=M_{0,1}=M_{-1,0}=M_{0,-1}
=\sqrt{2}m^2\nu\frac{\hbar^2k^2r^2}{m^2}\frac{
\sin\theta\cos\phi\sin^2\frac{\theta}{2}}
{\left(1+\nu^2\sin^2\frac{\theta}{2}\right)^2}+O(\hbar^3)
\nonumber\\
&&M_{1,1}=M_{-1,-1}=M_{-1,1}=M_{1,-1}=2m^2\nu^2
\nonumber\\
&&M_{2,0}=M_{0,2}=M_{-2,0}=M_{0,-2}=
-m^2\nu^2-m^2\frac{\hbar^2k^2r^2}{m^2}\frac{\nu^2\sin^4\frac{\theta}{2}}
{\left(1+\nu^2\sin^2\frac{\theta}{2}\right)}+O(\hbar^3)
\nonumber
\end{eqnarray}
for (A),
\begin{eqnarray}
&& M_{0,0}=-4m^2(1+\nu^2)-4m^2\frac{\hbar^2 k^2 r^2}{m^2}
\frac{n^2\sin^2\frac{\theta}{2}}{n(n-\sin\theta\cos\phi)
+\frac{m^2}{p_\rmi^2}\nu^2\sin^2\frac{\theta}{2}}+O(\hbar^3)\nonumber\\
&&M_{1,0}=M_{0,1}=M_{-1,0}=M_{0,-1}\nonumber\\\nonumber\\
&&=\sqrt{2}m^2\frac{\hbar^2k^2r^2}{m^2}
\frac{n^3\nu\frac{m}{p_{\rmi}}\sin^2\frac{\theta}{2}
\left(2\sin^2\frac{\theta}{2}-\sin\theta\cos\phi\right)}
{\left(n(n-\sin\theta\cos\phi)+\frac{m^2}{p^2_{\rmi}}
\nu^2\sin^2\frac{\theta}{2}\right)^2}+O(\hbar^3)
\nonumber\\
&&M_{1,1}=M_{-1,-1}=M_{-1,1}=M_{1,-1}=2m^2\nu^2
\nonumber\\
&&M_{2,0}=M_{0,2}=M_{-2,0}=M_{0,-2}=
\nonumber\\
&&=
-m^2\nu^2-m^2\frac{\hbar^2k^2r^2}{m^2}
\frac{n^2\nu^2\sin^4\frac{\theta}{2}}
{\left(n(n-\sin\theta\cos\phi)+
\frac{m^2}{p_\rmi^2}\nu^2\sin^2\frac{\theta}{2}\right)^2}+O(\hbar^3)
\nonumber
\end{eqnarray}
for (B), and 
\begin{eqnarray}
&&M_{0,0}=-4m^2(1+\nu^2)-4m^2\frac{\hbar^2 k^2 r^2}{m^2}
\sin^2\frac{\theta}{2}+O(\hbar^3)\nonumber\\
&&M_{1,0}=M_{0,1}=M_{-1,0}=M_{0,-1}
=\sqrt{2}m^2\nu\frac{\hbar^2k^2r^2}{m^2}
\sin\theta\cos\phi\sin^2\frac{\theta}{2}+O(\hbar^3)\nonumber\\
&&M_{1,1}=M_{-1,-1}=M_{-1,1}=M_{1,-1}=2m^2\nu^2
\nonumber\\
&&M_{2,0}=M_{0,2}=M_{-2,0}=M_{0,-2}=
-m^2\nu^2-m^2\frac{\hbar^2k^2r^2}{m^2}\nu^2\sin^4\frac{\theta}{2}
+O(\hbar^3),
\nonumber
\end{eqnarray}
for (C), respectively.
Thus, the quantum correction in the kernel $M_{i,j}$
arises at the order of $\hbar^2$.  

\end{appendix}


\begin{thebibliography}{0}
\bibitem{Higuchi} A. Higuchi and P. J. Walker, Phys. Rev. D {\bf 80} 105019 (2009).
\bibitem{YN} K. Yamamoto, G. Nakamura, Phys. Rev. D {\bf 83} 045030 (2011).
\bibitem{KNY}R. Kimura, G. Nakamura, K. Yamamoto Phys. Rev. D {\bf 83} 045015 (2011).
\bibitem{NSY} H. Nomura, M. Sasai, K. Yamamoto, J. Cosmol. Astropart. Phys. 
{\bf 11} (2006) 013.
\bibitem{HM1}A. Higuchi and G. D. R. Martin, Found. Phys. {\bf 35} 1149 (2005).
\bibitem{HM2}A. Higuchi and G. D. R. Martin, Phys. Rev. D {\bf 73} 025019 (2006).
\bibitem{HM3}A. Higuchi and G. D. R. Martin, Phys. Rev. D {\bf 74} 125002 (2006).
\bibitem{Jackson} J. D. Jackson, {\it Classical Electrodynamics} (Wiley, 1998).
\bibitem{Volkov}D. Volkov, Z. Phys. {\bf 94}, 250 (1935).
\bibitem{NikishovRitus} A. I. Nikishov and V. I. Ritus Sov. Phys. JETP {\bf 19}, 529 (1964)
\bibitem{BK}L. S. Brown and T. W. B. Kibble, Phys. Rev. {\bf 133} A705 (1964).
\bibitem{Brown} L. S. Brown, Phys. Rev. {\bf 138} B740 (1965).
\bibitem{Wein} S. Weinberg, Phys. Rev. D {\bf 72} 043514 (2005).
\bibitem{Schwinger} J. Schwinger, Phys. Rev. {\bf 82}, 664 (1951)
\bibitem{Nikishov} A. I. Nikishov, Sov. Phys. JETP {\bf 32}, 690 (1971)
\bibitem{DWB} B. S. De~Witt and R. W. Brehme, Ann. Phys. (N.Y) {\bf 9}, 220 (1960)
\bibitem{Hobbsa} J. M. Hobbs, Ann. Pys. (N.Y.) {\bf 47}, 141 (1968)
\bibitem{Hobbsb} J. M. Hobbs, Ann. Pys. (N.Y.) {\bf 47}, 166 (1968)
\bibitem{Futamase}T. Futamase, et~al., Prog. Theor. Phys. {\bf 96}, 113 (1996).
\bibitem{HWU} A. Higuchi and P. J. Walker, Phys. Rev. D {\bf 79} 105023 (2009).
\bibitem{AEL} P. Adshead, R. Easther and E. A. Lim, 
Phys. Rev. D {\bf 80} 083521 (2009).
\bibitem{CT} P. Chen, T. Tajima, Phys. Rev. Lett {\bf 83} 256 (1999).
\bibitem{Iso} S. Iso, Y. Yamamoto, S. Zhang, Phys. Rev. D {\bf 84} 025005 (2011). 
\end{thebibliography}
\end{document}